# BAM: A Balanced Attention Mechanism for Single Image Super Resolution

**AAAI Press (This is the Author Name style)**
Association for the Advancement of Artificial Intelligence (This is the Affiliation and Address style)
pubforms22@aaai.org (This is also the Affiliation and Address style)

**Abstract**

Recovering texture information from the aliasing regions has always been a major challenge for Single Image Super Resolution (SISR) task. These regions are often submerged in noise so that we have to restore texture details while suppressing noise. To address this issue, we propose a Balanced Attention Mechanism (BAM), which consists of Avgpool Channel Attention Module (ACAM) and Maxpool Spatial Attention Module (MSAM) in parallel. ACAM is designed to suppress extreme noise in the large scale feature maps while MSAM preserves high-frequency texture details. Thanks to the parallel structure, these two modules not only conduct self-optimization, but also mutual optimization to obtain the balance of noise reduction and high-frequency texture restoration during the back propagation process, and the parallel structure makes the inference faster. To verify the effectiveness and robustness of BAM, we applied it to 10 SOTA SISR networks. The results demonstrate that BAM can efficiently improve the networks' performance, and for those originally with attention mechanism, the substitution with BAM further reduces the amount of parameters and increases the inference speed. Moreover, we present a dataset with rich texture aliasing regions in real scenes, named realSR7. Experiments prove that BAM achieves better super-resolution results on the aliasing area.

## Introduction

Single image super-resolution (SISR) is one of the popular computer vision research topics (Wang et al. 2020; Anwar et al. 2020), which aims to reconstruct a high-resolution (HR) image from a low-resolution (LR) image. With the success of deep learning prevailed in computer vision, many convolutional neural network (CNN) based super-resolution (SR) methods have been proposed. According to their architectures, they can be categorized into linear (Zhang et al. 2017a; Zhang et al. 2017b; Dong et al. 2015; Shi et al. 2016; Dong et al. 2016), recursive (Tai et al. 2017a; Tai et al. 2017b; Kim et al. 2016b), densely connected (Abbass et al. 2020; et al. 2016; Kim et al. 2016a), residual (Jiao et al. 2020; Fan Haris et al. 2018; Tong et al. 2017), multi-path (Park et al. 2018), and adversarial (Wang et al. 2018) designs. In order to further improve the quality of SR results while controlling parameter amounts, attention mechanisms were adopted in some SISR networks. At the same time, there exist quite a lot of excellent SISR networks (EDSR(Lim et al. 2017), CARN (Ahn et al. 2018), MSRN(Qin et al. 2020), s-LWSR(Li et al. 2020), AWSRN(Wang et al. 2019)) without the attention mechanism. One motivation of our work is to propose a plug-and-play attention mechanism for them so that their applications can be more extensive, and make it more fair to compare these networks with those with attention RCAN(Zhang et al. 2018), IMDN(Hui et al. 2019), PAN(Zhao et al. 2020) and DRLN(Anwar et al. 2020). The attention mechanism CBAM(Woo et al.2018) and SE(Hu et al. 2018) were first applied to classification tasks. Due to its remarkable results in classification, researchers have made great efforts along this direction and expanded its application to SISR tasks. However, the SISR networks are so diverse that the attention module is usually designed solely for a specific network structure. These separately proposed attention mechanisms require a baseline to compare with in order to verify their effectiveness. Therefore, another motivation of our work is to provide a baseline of attention mechanism for SISR. Actually, our proposed BAM is not only more efficient but also more lightweight than the attention mechanisms proposed in RCAN, IMDN, PAN and DRLN, which has been proved in our experiments. Last but not lease, one major problem for the existing SISR networks is the information restoration in the texture aliasing area, so our biggest motivation is to overcome this problem by designing a specific attention mechanism.

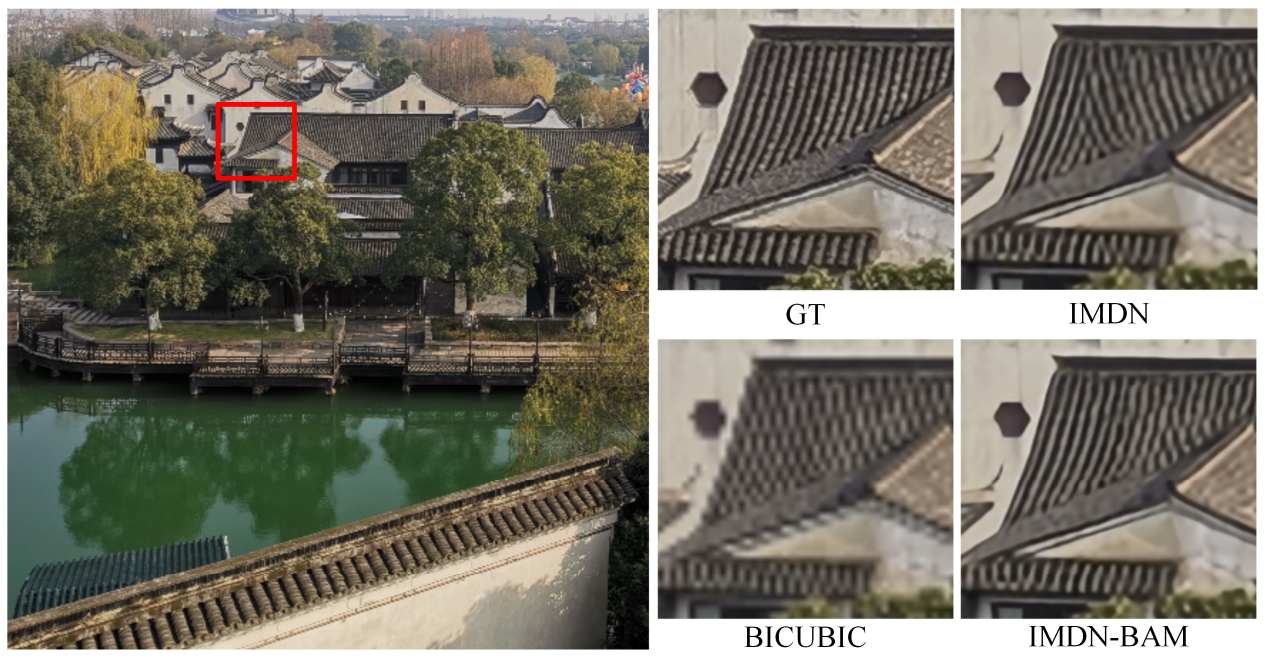

Figure 1: Comparison of ×4 SR results of IMDN and IMDN-BAM on the realSR7 dataset. IMDN-BAM shows better super-resolution results on texture aliasing areas.

For the SISR networks without attention, BAM can be easily inserted behind the basic block or before the up-sampling layer. And for those with attention, BAM can seamlessly replace their original attention mechanism. We experimented on 6 networks without attention and 4 with attention to verify the effectiveness and robustness of BAM. Our contributions are summarized as follows:

- We propose a lightweight and efficient attention mechanism, BAM, for the SISR task. BAM can restore high-frequency texture information as much as possible while suppressing the extreme noise in the large scale feature maps. Furthermore, the parallel structure can improve the inference speed.
- We conduct comparative experiments on 10 SOTA SISR networks. The insertion or replacement of BAM generally improves the PSNR and SSIM (Wang et al. 2004) values of the SR results and the visual quality with less training data, and for those with attention, the replacement of BAM further reduces the amount of parameters and accelerates the inference speed. What's more, for lightweight SISR networks, the comparative experiments illustrate that BAM can generally improve their performance but barely increase or even decrease the parameters, which is significant for their deployment on terminals.
- We present a real-scene SISR dataset realSR7 considering the practical texture aliasing issue. BAM can achieve better SR performance on this realistic dataset.

Our codes, pre-trained models and the realSR7 dataset are available at: https://github.com/dandingbudanding/BAM.

## Related works

In this section, we will introduce the 10 SISR networks used in our control experiments.

### SISR networks without attention

EDSR (Lim et al. 2017) removes the BN layer and the last activation layer in the residual network. Our BAM module is inserted before the up-sampling layer. To achieve real-time performance, Namhyuk Ahn proposed CARN (Ahn et al. 2018) in which the local and global cascade structures can integrate features from multiple layers, which enables learning multi-scale information of the feature maps. Its lightweight variant, CARN-M, compromises the performance for speed. For these two networks, BAM is inserted behind each CARN block. MSRN (Qin et al. 2020) combines local multi-scale features with global features to fully exploit the LR image, which solves the issue of feature disappearance during propagation. BAM will be concatenated to the end of each MSRN block.

s-LWSR (Li et al. 2020) applies the encoder-decoder structure for the SISR problem. In order to adapt to different scenarios, three networks of different size, s-$LWSR_{16}$, s-$LWSR_{32}$ and s-$LWSR_{64}$, were proposed. Here we choose the middle-size one, s-$LWSR_{32}$. For s-$LWSR_{32}$, the BAM will be inserted before the up-sampling layer. A novel local fusion block is designed in AWSRN (Wang et al. 2019) for efficient residual learning, which consists of stacked adaptive weighted residual units and a local residual fusion unit. It can achieve efficient flow and fusion of information and gradients. Moreover, an adaptive weighted multi-scale (AWMS) module is proposed to not only make full use of the features in reconstruction layer but also reduce the amount of parameters by analyzing the information redundancy between branches of different scales. Different from the aforementioned networks, BAM will be inserted before the AWMS module.

### SISR networks with attention

RCAN (Zhang et al. 2018) utilized a residual-in-residual (RIR) structure to construct the whole network, which allows the rich low-frequency information to directly propagate to the rear part through multiple skip connections. Thus the network can focus on learning high-frequency information. What's more, a channel attention (CA) mechanism was utilized to adaptively adjust features by considering the interdependence between channels. IMDN (Hui et al. 2019) is a representative lightweight SISR network with attention mechanism. It is constructed by the cascaded information multi-distillation blocks (IMDB) consisting of distillation and selective fusion parts. The distillation module extracts hierarchical features step-by-step, and fusion module aggregates them according to the importance of candidate features, which is evaluated by the proposed contrast-aware channel attention (CCA) mechanism. PAN (Zhao et al. 2020) is the winning solution of AIM2020 VTSR Challenge. It proposed a pixel attention (PA) mechanism, similar to channel attention and spatial attention. The difference is that PA generates 3D attention maps, which allows the performance improvement with fewer parameters. DRLN (Anwar et al. 2020) employs cascading residual on the residual structure to allow the flow of low-frequency information so that the network can focus on learning high and mid-level features. Moreover, it proposes a Laplacian attention (LA) to model the crucial features to learn the inter-level and intra-level dependencies between the feature maps. In the comparative experiments, CA, CCA, PA, and LA will be replaced with BAM.

## Proposed method

Some texture details in low-resolution images are often overwhelmed by extreme noises, which leads to a major difficulty to recover texture information from the texture aliasing area. To solve this problem, we proposed the BAM composed of ACAM and MSAM in parallel, where ACAM is dedicated to suppressing extreme noise in the large scale feature maps and MSAM tries to pay more attention to the high-frequency texture details. Moreover, the parallel structure of BAM will allow not only self-optimization, but also mutual optimization of the channel and spatial attention during the gradient backpropagation process so as to achieve a balance between them. It can obtain the best noise reduction and high-frequency information recovery capabilities, and the parallel structure can speed up the inference process. The schematic of BAM is shown in Figure 2. Since ACAM and MSAM generate vertical and horizontal attention weights for the input feature maps respectively, the dimension of their output is inconsistent. One is $N \times C \times 1 \times 1$ and the other is $N \times 1 \times H \times W$. Thus, we use broadcast-multiplication to fuse them into an $N \times C \times H \times W$ weight tensor, and then multiply it with the input feature maps element-wisely. Here, N is the batch size (N=16 in our experiments), C is the number of channels of the feature maps, H and W are the height and width of the feature maps. In ACAM, avgpool operation is used to obtain the average value of each feature map, while in MSAM, maxpool operation is used to get the max value among the C channels for each position on the feature map, and they can be expressed as

$$Avgpool(N,C,1,1) = \frac{1}{H \times W} \sum_{h=0}^{H-1} \sum_{w=0}^{W-1} F(N,C,h,w), \quad (1)$$

$$Maxpool(N,1,H,W) = \max\{F(N,c,H,W), c \in [0, C-1]\}, \quad (2)$$

where $F \in \mathbb{R}^{N \times C \times H \times W}$ represents the input feature maps, $\max\{\}$ means to get the max value.

## ACAM

Channel attention needs to find channels with more important information from the input feature maps and give them higher weights. It is highly likely for a channel with the dimension of $H \times W$ (in our experiments, $H = W \geq 64$) to contain some abnormal extrema. Maxpool will pick these extreme values as noise and get the wrong attention information, which will make the texture recovery more difficult. Therefore, we only use avgpool to extract channel information so that it complies with Occam's razor principle when suppressing extreme noise and then pass it through a multi-layer perceptron (MLP) composed of two point-wise convolution layers. To increase the nonlinearity of MLP, PReLU (He et al. 2016) is used to activate the first convolution layer output. In addition, to reduce the parameter amount and computational complexity of ACAM, MLP adopts the bottleneck architecture (He et al.2016). The number of input channels is r times the number of output channels for the first convolution layer. After PReLU activation, the number of channels is restored by the second convolution layer. Finally, the channel weights are generated by a sigmoid activation function. The generation process of ACAM can be described by

$$ACAM(F) = Sigmiod[\mathcal{F}_{n/r->n}^{k \times k}(PReLU(\mathcal{F}_{n->n/r}^{k \times k}(Avgpool(F))))], \quad (3)$$

where $\mathcal{F}_{n->n/r}^{k \times k}$ represents the convolution layer with the kernel size of $k \times k$ (for Eq.3, $k=1$), the input channel number of $n$ and the output channel number of $n/r$, r is set to 16 and n is determined by the channel numbers of the input feature maps in experiments.

## MSAM

Spatial attention generates weights for the horizontal section of the input feature maps. Its goal is to find lateral areas which contribute most to the final HR reconstruction and give them higher weights. These areas usually contain high-frequency details in the form of extreme values in the channel. Thus, using maxpool operation for spatial attention is appropriate. The output of maxpool passes a convolution layer with large receptive field of $k \times k$ (for Eq.4, $k=7$), and then gets activated by the sigmoid function to obtain the

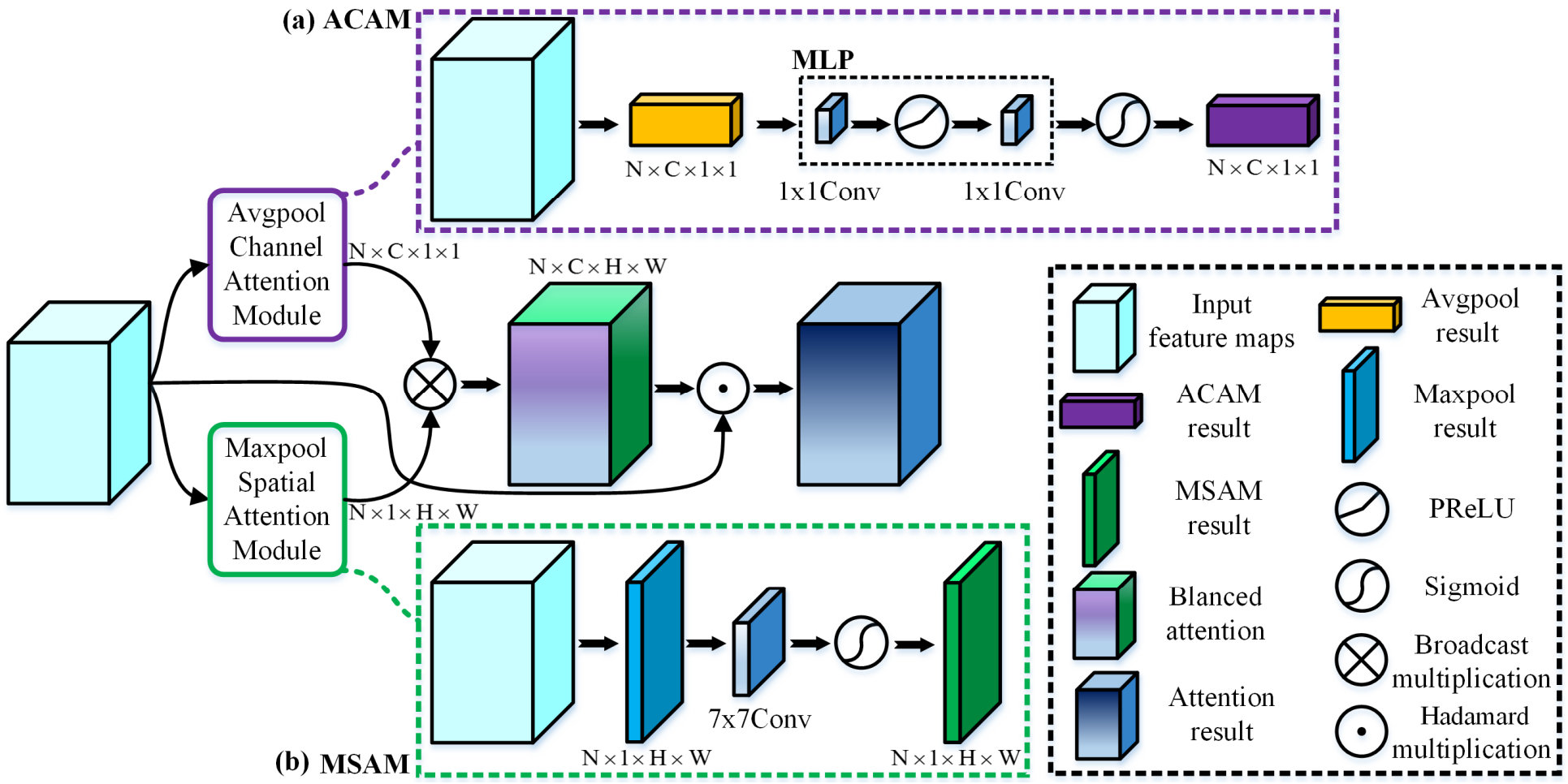

Figure 2: BAM. The channel attention from ACAM and the spatial attention from MSAM are fused by broadcast multiplication and then multiplied with the input feature maps element-wisely to obtain the final attention result. (a) ACAM. The channel attention information is extracted by avgpool. (b) MSAM. The spatial attention information is extracted by maxpool.

spatial attention weights. This design effectively controls the amount of parameters. It can be expressed by

$$MSAM(x) = Sigmiod[\mathcal{F}_{1->1}^{7\times7}(Maxpool(x))] . \quad (4)$$

### BAM

There are two innovations in the design of BAM. One is that the ACAM tries to suppress the extreme noise and the MSAM tries to maintain the texture information. The other is the parallel structure, which makes the generation process of channel attention and spatial attention independent of each other and allows the mutual optimization of two attentions during the backpropagation. The combination of these two innovations enables BAM to recover as much high-frequency information as possible from the texture aliasing area. Ablation experiments prove that the current design of BAM can effectively control the parameter amount and obtain better performance than the original networks, evaluated by PSNR and SSIM metrics. The formula of BAM is

$$BAM(F) = ACAM(F) \otimes MSAM(F) \odot F , \quad (5)$$

where $\otimes$ means broadcast multiplication and $\odot$ stands for Hadamard multiplication. Because the outputs of ACAM and MSAM have different dimensions, we utilize broadcast multiplication to fuse them and then element-wisely multiply it with the input feature maps to obtain the final attention results. ACAM and MSAM are self-optimized in their respective gradient backpropagation process. To reveal the mutual optimization of ACAM and MSAM in the gradient backpropagation process of BAM, we give the partial derivative of BAM concerning the input feature maps $F$ as follows:

$$\frac{\partial BAM(F)}{\partial F} = \frac{\partial ACAM(F)}{\partial F} \otimes MSAM(F) \odot F + ACAM(F) \otimes \frac{\partial MSAM(F)}{\partial F} \odot F + ACAM(F) \otimes MSAM(F) . \quad (6)$$

As illustrated in Eq.6, not only is ACAM and MSAM related to each other but also related to each other's first-order partial differentials(The gradient), which means ACAM and MSAM can optimize mutually in the gradient backpropagation process of BAM.

## Experiments and discussions

### Datasets and metrics

The training sets are various for different SISR networks, and for the deep learning task, the richer the data is, the better the results would be. To verify the efficient performance of the proposed BAM, following AWSRN, RCAN and IMDN, we use 800 high-quality (2K resolution) images from DIV2K (Agustsson et al. 2017) as the training set, and evaluate on Set5 (Bevilacqua et al. 2012), Set14 (Zeyde et al. 2010), BSD100 (Martin et al. 2001), and Manga109 (Narita et al. 2017) with the PSNR and SSIM metrics under the upscaling factors of ×2, ×3, and ×4 respectively, for ablation experiments we add Urban100 (Huang et al. 2015) and our realSR7 for validation. In all the experiments, bicubic interpolation is utilized as the resizing method. We calculate the metrics on the luminance channel (Y channel of the YCbCr channels converted from the RGB).

### Implementation details

During the training, we use the RGB patches with size of 64×64 from the LR input together with its corresponding HR patches. We only apply data augmentation to the training data. Specifically, the 800 image pairs in training set are cropped into five pairs from the four corners and center of the original image so that the training set is expanded by 5 times to 4000 image pairs. In addition, we randomly rotate and flip them during the training process. For optimization, Adam is used and its initial learning rate is set as 0.0001, which will be halved at every 200 epochs. The batch size is set as 16. We train for a total of 1000 epochs. The loss function for training is L1 loss function. We adopt pytorch1.1.0 framework to implement experiments on the desktop computer with 3.4 GHz Intel Xeon-E5-2643-v3 CPU, 64G RAM, and two NVIDIA GTX 1080Ti GPUs.

### Comparison experiments

For the convenience of discussion, we refer to the original networks as the control group, the BAM versions as the experimental group and add the "BAM" suffix to the networks name. The control experiments' results are summarized in Table 1. It can be seen that, for the three scale factors, the highest PSNR and SSIM metrics are all achieved by DRLN-BAM, and for ×4 up-sampling scale, the PSNR/SSIM metrics improvements on four benchmarks are{0.03/0.0007, 0.17/0.0036, 0.95/0.0289, 0.55/0.0070} separately, meanwhile the reduction of the parameter amount is 266.7K. Compared with the original attention mechanism of DRLN, BAM reduces the parameters, but obtains better performance. Actually, some control experiments used additional data sets for training in their original papers. Although our experimental groups have the disadvantage of a smaller training set, but can achieve a better results than the corresponding control groups. For lightweight networks such as PAN and IMDN, it is traditionally quite difficult to further improve their performance. The proposal of BAM makes it possible to enhance them even with reduced parameters, which is of great significance for their deployment in realistic cases. The results of the comparative experiments in Table 1 show that for the networks without attention, the incorporation of BAM can further increase their performance indicated by PSNR and SSIM metrics by only adding a small number of parameters.

Figure 3 displays the visual perception comparison between the ×4 SR results of the experimental group and the

Table 1: Control experiment results on 10 SISR networks, the lightweight SISR networks are marked in **bold black**. The parameter amount is calculated based on a 240×360 RGB image. The growth or decline of PSNR/SSIM compared with the corresponding control group is indicated by ↑ and ↓ respectively (**the higher the better**). The best two results are highlighted in red and blue colors respectively.

| Scale | Method | Param | Set5 PSNR/SSIM | Set14 PSNR/SSIM | BSD100 PSNR/SSIM | Manga109 PSNR/SSIM |
|---|---|---|---|---|---|---|
| ×2 | EDSR(CVPRW'17) | 40729.6K | 38.11/0.9601 | 33.92/0.9195 | 32.32/0.9013 | - |
| | EDSR-BAM | 40737.9K | 38.19/0.9613 ↑0.08/↑0.0012 | 34.00/0.9213 ↑0.08/↑0.0018 | 34.20/0.9273 ↑1.88/↑0.0260 | 39.72/0.9806 |
| | **CARN**(ECCV'18) | 1592.0K | 37.76/0.9590 | 33.52/0.9166 | 32.09/0.8978 | - |
| | CARN-BAM | 1593.7K | 37.84/0.9600 ↑0.08/↑0.0010 | 33.55/0.9167 ↑0.03/↑0.0001 | 33.90/0.9245 ↑1.81/↑0.0267 | 38.68/0.9787 |
| | **CARN-M**(ECCV'18) | 1161.3K | 37.53/0.9583 | 33.26/0.9141 | 31.92/0.8960 | - |
| | CARN-M-BAM | 1163.0K | 37.75/0.9597 ↑0.22/↑0.0014 | 33.44/0.9158 ↑0.18/↑0.0017 | 33.81/0.9237 ↑1.89/↑0.0277 | 38.48/0.9783 |
| | MSRN(ECCV'18) | 5930.3K | 38.08/0.9605 | 33.74/0.9170 | 32.23/0.9013 | 38.64/0.9771 |
| | MSRN-BAM | 5934.9K | 38.11/0.9610 ↑0.03/↑0.0005 | 33.84/0.9192 ↑0.10/↑0.0018 | 34.12/0.9265 ↑1.89/↑0.0252 | 39.45/0.9801 ↑0.81/↑0.0030 |
| | **s-LWSR$_{32}$**(TIP'19) | 534.1K | - | - | - | - |
| | s-LWSR$_{32}$-BAM | 534.3K | 37.91/0.9603 | 33.63/0.9174 | 33.97/0.9252 | 38.82/0.9791 |
| | **AWSRN**(CVPR'19) | 1396.9K | 38.11/0.9608 | 33.78/0.9189 | 32.26/0.9006 | 38.87/0.9776 |
| | AWSRN-BAM | 1397.2K | 38.14/0.9610 ↑0.03/↑0.0002 | 33.91/0.9201 ↑0.13/↑0.0012 | 34.15/0.9268 ↑1.89/↑0.0262 | 39.41/0.9802 ↑0.54/↑0.0026 |
| | RCAN(ECCV'18) | 15444.7K | 38.27/0.9617 | 34.23/0.9225 | 32.46/0.9031 | 39.44/0.9786 |
| | RCAN-BAM | 15441.7K | 38.32/0.9618 ↑0.05/↑0.0001 | 34.25/0.9230 ↑0.02/↑0.0005 | 34.29/0.9282 ↑1.83/↑0.0251 | 39.86/0.9806 ↑0.42/↑0.0020 |
| | **IMDN**(ACM MM'19) | 694.4K | 38.00/0.9605 | 33.63/0.9177 | 32.19/0.8996 | 38.88/0.9774 |
| | IMDN-BAM | 694.3K | 38.03/0.9607 ↑0.03/↑0.0002 | 33.73/0.9183 ↑0.10/↑0.0006 | 34.05/0.9259 ↑1.86/↑0.0263 | 39.33/0.9800 ↑0.45/↑0.0026 |
| | **PAN**(ECCVW'20) | 261.4K | 38.00/0.9605 | 33.59/0.9181 | 32.18/0.8997 | 38.70/0.9773 |
| | PAN-BAM | 261.0K | 38.00/0.9606 ↑0.00/↑0.0001 | 33.70/0.9181 ↑0.11/↑0.0000 | 34.03/0.9255 ↑1.85/↑0.0258 | 39.19/0.9797 ↑0.31/↑0.0024 |
| | DRLN(TPAMI'20) | 34430.2K | 38.27/0.9616 | 34.28/0.9231 | 32.44/0.9028 | 39.58/0.9786 |
| | DRLN-BAM | 34163.4K | 38.32/0.9619 ↑0.05/↑0.0003 | 34.42/0.9237 ↑0.14/↑0.0006 | 34.33/0.9284 ↑1.89/↑0.0256 | 40.41/0.9820 ↑0.83/↑0.0034 |
| ×3 | EDSR(CVPRW'17) | 43680.0K | 34.65/0.9282 | 30.52/0.8462 | 29.25/0.8091 | - |
| | EDSR-BAM | 43688.3K | 35.26/0.9417 ↑0.61/↑0.0135 | 31.15/0.8607 ↑0.63/↑0.0145 | 29.73/0.8212 ↑0.48/↑0.0121 | 34.04/0.9495 |
| | **CARN**(ECCV'18) | 1592.0K | 34.29/0.9255 | 30.29/0.8407 | 29.06/0.8034 | - |
| | CARN-BAM | 1593.7K | 34.93/0.9392 ↑0.64/↑0.0137 | 30.93/0.8560 ↑0.64/↑0.0153 | 29.57/0.8171 ↑0.51/↑0.0137 | 33.52/0.9456 |
| | **CARN-M**(ECCV'18) | 1161.3K | 33.99/0.9236 | 30.08/0.8367 | 28.91/0.8000 | - |
| | CARN-M-BAM | 1163.0K | 34.81/0.9383 ↑0.82/↑0.0147 | 30.84/0.8540 ↑0.76/↑0.0173 | 29.49/0.8150 ↑0.58/↑0.0150 | 33.31/0.9438 |
| | MSRN(ECCV'18) | 6115.0K | 34.38/0.9262 | 30.34/0.8395 | 29.08/0.8041 | 33.44/0.9427 |
| | MSRN-BAM | 6119.5K | 35.20/0.9412 ↑0.82/↑0.0150 | 31.10/0.8590 ↑0.76/↑0.0195 | 29.66/0.8195 ↑0.58/↑0.0154 | 33.90/0.9483 ↑0.46/↑0.0056 |
| | **s-LWSR$_{32}$**(TIP'19) | 580.4K | - | - | - | - |
| | s-LWSR$_{32}$-BAM | 580.6K | 34.98/0.9395 | 30.94/0.8569 | 29.58/0.8175 | 33.50/0.9459 |
| | **AWSRN**(CVPR'19) | 1476.1K | 34.52/0.9281 | 30.38/0.8426 | 29.16/0.8069 | 33.85/0.9463 |
| | AWSRN-BAM | 1476.5K | 35.13/0.9408 ↑0.61/↑0.0127 | 31.09/0.8590 ↑0.71/↑0.0164 | 29.65/0.8191 ↑0.49/↑0.0132 | 33.82/0.9478 ↑0.03/↑0.0015 |
| | RCAN(ECCV'18) | 15629.3K | 34.74/0.9299 | 30.65/0.8482 | 29.32/0.8111 | 34.44/0.9499 |
| | RCAN-BAM | 15626.3K | 35.36/0.9424 ↑0.62/↑0.0125 | 31.22/0.8611 ↑0.57/↑0.0129 | 29.75/0.8215 ↑0.43/↑0.0104 | 34.07/0.9501 ↓0.37/↑0.0002 |
| | **IMDN**(ACM MM'19) | 703.1K | 34.36/0.9270 | 30.32/0.8417 | 29.09/0.8046 | 33.61/0.9445 |
| | IMDN-BAM | 703.0K | 35.06/0.9405 ↑0.70/↑0.0135 | 30.99/0.8568 ↑0.67/↑0.0151 | 29.61/0.8181 ↑0.52/↑0.0135 | 33.80/0.9474 ↑0.19/↑0.0029 |
| | **PAN**(ECCVW'20) | 261.4K | 34.40/0.9271 | 30.36/0.8423 | 29.11/0.8050 | 33.61/0.9448 |
| | PAN-BAM | 261.0K | 34.77/0.9379 ↑0.37/↑0.108 | 30.88/0.8545 ↑0.52/↑0.0122 | 29.50/0.8145 ↑0.39/↑0.0095 | 33.19/0.9435 ↓0.42/↓0.0013 |
| | DRLN(TPAMI'20) | 34614.8K | 34.78/0.9303 | 30.73/0.8488 | 29.36/0.8117 | 34.71/0.9509 |
| | DRLN-BAM | 34348.1K | 35.42/0.9431 ↑0.64/↑0.0128 | 31.32/0.8628 ↑0.59/↑0.0140 | 29.81/0.8224 ↑0.45/↑0.0107 | 34.73/0.9527 ↑0.02/↑0.0018 |
| ×4 | EDSR(CVPRW'17) | 43089.9K | 32.46/0.8968 | 28.80/0.7876 | 27.71/0.7420 | - |
| | EDSR-BAM | 43098.2K | 32.46/0.8986 ↑0.00/↑0.0018 | 28.92/0.7901 ↑0.12/↑0.0025 | 28.63/0.7688 ↑0.92/↑0.0268 | 31.49/0.9219 |
| | **CARN**(ECCV'18) | 1592.0K | 32.13/0.8940 | 28.60/0.7810 | 27.58/0.7350 | - |
| | CARN-BAM | 1593.7K | 32.17/0.8944 ↑0.04/↑0.0004 | 28.72/0.7839 ↑0.12/↑0.0029 | 28.46/0.7628 ↑0.88/↑0.0278 | 30.81/0.9140 |
| | **CARN-M**(ECCV'18) | 1161.3K | 31.92/0.8900 | 28.42/0.7760 | 27.44/0.7300 | - |
| | CARN-M-BAM | 1163.0K | 31.98/0.8915 ↑0.06/↑0.0015 | 28.54/0.7792 ↑0.08/↑0.0032 | 28.35/0.7593 ↑0.91/↑0.0293 | 30.44/0.9091 |
| | MSRN(ECCV'18) | 6082.6K | 32.07/0.8903 | 28.60/0.7751 | 27.52/0.7273 | 30.17/0.9034 |
| | MSRN-BAM | 6078.0K | 32.14/0.8940 ↑0.07/↑0.0037 | 28.66/0.7830 ↑0.06/↑0.0079 | 28.45/0.7626 ↑0.93/↑0.0353 | 30.69/0.9122 ↑0.52/↑0.0088 |
| | **s-LWSR$_{32}$**(TIP'19) | 571.1K | 32.04/0.8930 | 28.15/0.7760 | 27.52/0.7340 | - |
| | s-LWSR$_{32}$-BAM | 571.3K | 32.07/0.8935 ↑0.03/↑0.0005 | 28.70/0.7843 ↑0.55/↑0.0083 | 28.48/0.7636 ↑0.96/↑0.0296 | 30.82/0.9137 |
| | **AWSRN**(CVPR2019) | 1587.1K | 32.27/0.8960 | 28.69/0.7843 | 27.64/0.7385 | 30.72/0.9109 |
| | AWSRN-BAM | 1587.4K | 32.29/0.8962 ↑0.02/↑0.0002 | 28.80/0.7863 ↑0.11/↑0.0020 | 28.54/0.7658 ↑0.90/↑0.0273 | 31.12/0.9172 ↑0.40/↑0.0063 |
| | RCAN(ECCV'18) | 15592.4K | 32.63/0.9002 | 28.87/0.7889 | 27.77/0.7436 | 31.22/0.9173 |
| | RCAN-BAM | 15589.4K | 32.64/0.9003 ↑0.01/↑0.0001 | 29.00/0.7918 ↑0.13/↑0.0029 | 28.69/0.7710 ↑0.92/↑0.0274 | 31.09/0.9209 ↑0.13/↑0.0036 |
| | **IMDN**(ACM MM'19) | 715.2K | 32.21/0.8948 | 28.58/0.7811 | 27.56/0.7353 | 30.47/0.9084 |
| | IMDN-BAM | 715.1K | 32.24/0.8955 ↑0.03/↑0.0007 | 28.75/0.7847 ↑0.17/↑0.0036 | 28.51/0.7642 ↑0.95/↑0.0289 | 31.02/0.9154 ↑0.55/↑0.0070 |
| | **PAN**(ECCVW'20) | 272.4K | 32.13/0.8948 | 28.61/0.7822 | 27.59/0.7363 | 30.51/0.9095 |
| | PAN-BAM | 271.6K | 32.14/0.8941 ↑0.01/↓0.0007 | 28.69/0.7831 ↑0.08/↑0.0009 | 28.46/0.7623 ↑0.87/↑0.0260 | 30.79/0.9131 ↑0.28/↑0.0036 |
| | DRLN(TPAMI'20) | 34577.9K | 32.63/0.9002 | 28.94/0.7900 | 27.83/0.7444 | 31.54/0.9196 |
| | DRLN-BAM | 34311.2K | 32.66/0.9005 ↑0.03/↑0.0003 | 29.08/0.7925 ↑0.06/↑0.0025 | 28.75/0.7714 ↑0.92/↑0.0270 | 31.90/0.9257 ↑0.36/↑0.0061 |

control group for IMDN and DRLN. IMDN and DRLN can represent the current lightweight and heavyweight top-level networks, respectively. As can be seen, the experimental group is capable of recovering more detailed information and has a significant improvement on the aliased texture areas, such as cloth textures and facial wrinkles. Whether for a lightweight network such as IMDN or a heavyweight network such as DRLN, the BAM replacement can further improve the visual quality of SR results with the reduced parameters. IMDN-BAM and DRLN-BAM can be utilized as baselines for the follow-up researches. Figure 4 illustrates the ×4 SR results of 5 groups of SISR networks with or without BAM on an image of BSD100 dataset. For the 3 networks without attention, EDSR, CARN and AWSRN, their BAM version only increases a few parameters but greatly improves the metrics. Especially for EDSR-BAM, which achieves a very obvious visual improvement compared to the control group. For the 2 lightweight networks with attention, IMDN and PAN, the BAM replacement increases the SR quality while reducing the number of parameters. For the 2 networks without attention, CARN and AWSRN, their BAM versions only increase a few parameters but greatly improves the metrics; for the two lightweight networks with attention, IMDN and PAN, the BAM replacement increases the SR quality while reducing the number of parameters.

### Ablation experiments

In order to verify the efficiency of BAM, we conduct ablation experiments on three scaling factors of ×2, ×3, and ×4 based on the IMDN. Its original attention module, CCA, is replaced with CA, SE, CBAM and BAM respectively. We evaluate on the five benchmarks of Set5, Set14, BSD100, Urban100 and Manga109 with PSNR and SSIM metrics. From the results of ablation experiments in Table 2, it can be found that under three scaling factors, all the networks using BAM obtain the highest SSIM and PSNR metrics on 5 benchmark datasets. Moreover, after replacing CCA with SE or CBAM, the performance of the model is worse than the original version, reflecting that the effective attention mechanism on classification tasks does not necessarily have the same effect on the SISR task. Moreover, Figure 5 shows the ×4 SR results of 5 attention mechanisms used in Table 2, where we can see that BAM maintains a great balance between noise suppression and high-frequency texture detail recovery. BAM is the best one to recover the texture aliasing area among the five attention mechanisms.

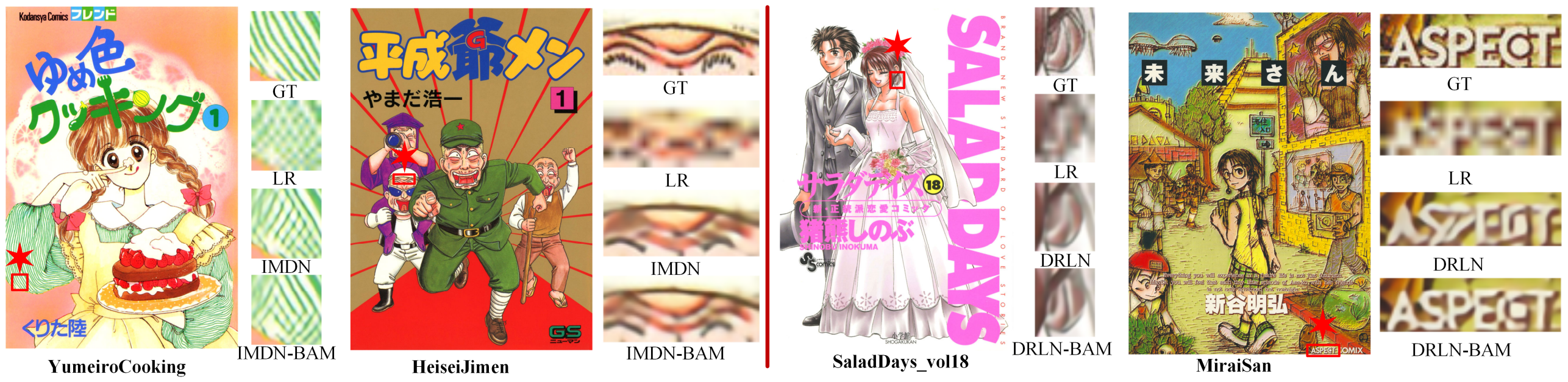

Figure 3: Visual perception comparison of the SR results from IMDN versus IMDN-BAM and DRLN versus DRLN-BAM on the Manga109 dataset under the scale factor of ×4.

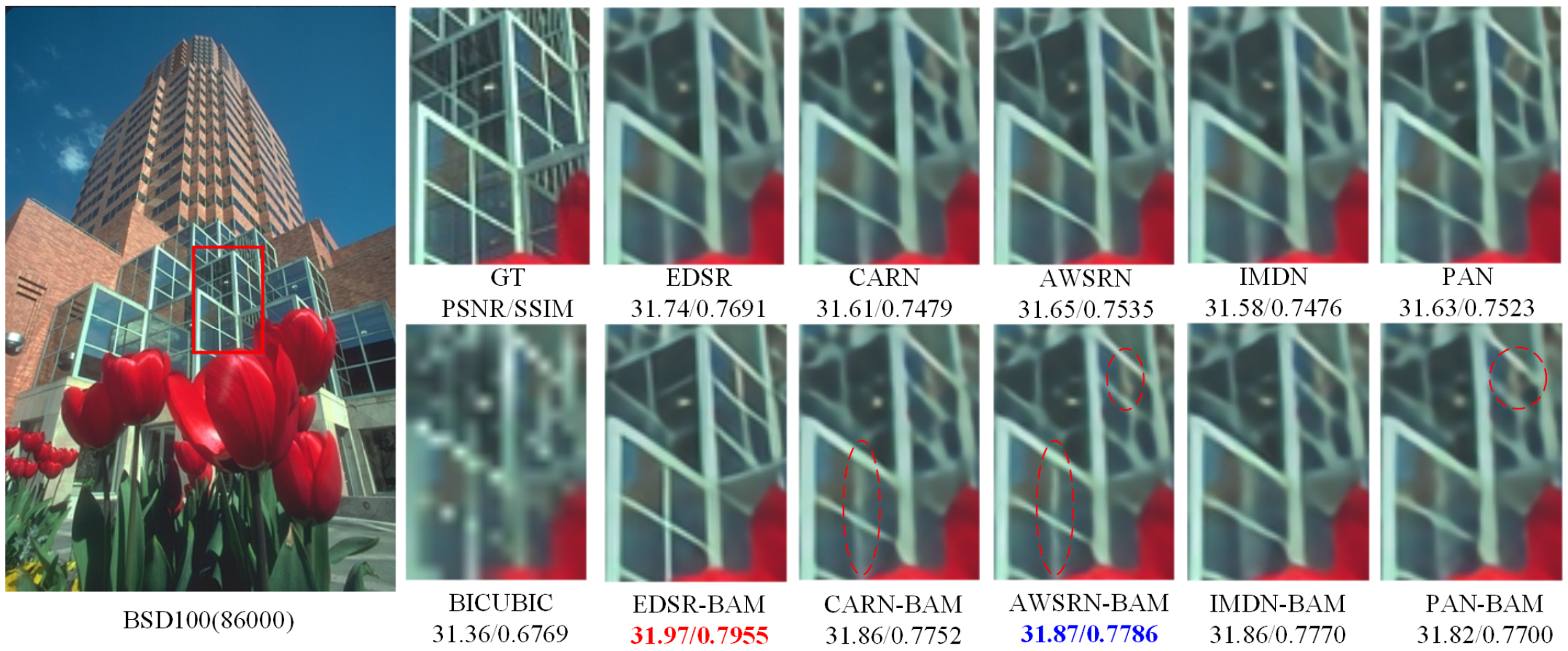

Figure 4: Metrics comparative experiments of five SISR networks under scaling factors of ×4. The best two results are highlighted in red and blue colors respectively. The red dashed ellipse is used to guide areas where the visual effect is not obvious improved. The improvement between EDSR-BAM and EDSR is significant.

Table 2: Ablation experiment results on Set5, Set14, BSD100, Urban100, Manga109 and realSR7, under three scaling factors of×2, ×3 and ×4 for IMDN with different attention mechanisms. The parameter amount and computational load are calculated based on an RGB image with the size of 240×360. The best two results are highlighted in red and blue colors respectively.

| Scale | Method | Param | GFLOPs | Set5 PSNR/SSIM | Set14 PSNR/SSIM | BSD100 PSNR/SSIM | Urban100 PSNR/SSIM | Manga109 PSNR/SSIM | realSR7 PSNR/SSIM |
|---|---|---|---|---|---|---|---|---|---|
| ×2 | IMDN(CCA) | 694.4K | 70.000 | 38.00/0.9605 | 33.63/0.9177 | 32.19/0.8996 | 32.17/0.9238 | 38.88/0.9774 | - |
| | IMDN(CA) | 694.4K | 70.000 | 37.86/0.9602 | 33.62/0.9173 | 33.94/0.9250 | 31.64/0.9234 | 38.97/0.9793 | - |
| | IMDN(SE) | 694.0K | 70.000 | 37.87/0.9602 | 33.60/0.9173 | 33.93/0.9249 | 31.69/0.9238 | 38.95/0.9792 | - |
| | IMDN(CBAM) | 694.6K | 70.086 | 37.87/0.9602 | 33.54/0.9168 | 33.89/0.9244 | 31.64/0.9234 | 38.62/0.9786 | - |
| | **IMDN(BAM)** | 694.3K | 70.027 | 38.03/0.9607 | 33.73/0.9183 | 34.05/0.9259 | 32.18/0.9283 | 39.33/0.9800 | - |
| ×3 | IMDN(CCA) | 703.1K | 70.831 | 34.36/0.9270 | 30.32/0.8417 | 29.09/0.8046 | 28.17/0.8519 | 33.61/0.9445 | - |
| | IMDN(CA) | 703.1K | 70.831 | 34.91/0.9392 | 30.91/0.8558 | 29.54/0.8168 | 28.92/0.8663 | 33.48/0.9456 | - |
| | IMDN(SE) | 702.7K | 70.831 | 34.93/0.9396 | 30.92/0.8558 | 29.54/0.8170 | 28.94/0.8667 | 33.53/0.9456 | - |
| | IMDN(CBAM) | 703.2K | 70.917 | 34.92/0.9393 | 30.91/0.8550 | 29.54/0.8164 | 28.82/0.8678 | 33.26/0.9444 | - |
| | **IMDN(BAM)** | 703.0K | 70.858 | 35.06/0.9405 | 30.99/0.8568 | 29.61/0.8181 | 29.11/0.8698 | 33.80/0.9474 | - |
| ×4 | IMDN(CCA) | 715.2K | 71.994 | 32.21/0.8948 | 28.58/0.7811 | 27.56/0.7353 | 26.04/0.7838 | 30.47/0.9084 | 30.29/0.8483 |
| | IMDN(CA) | 715.2K | 71.994 | 32.01/0.8921 | 28.59/0.7815 | 28.39/0.7611 | 25.74/0.7749 | 30.63/0.9111 | 30.32/0.8471 |
| | IMDN(SE) | 714.8K | 71.994 | 32.07/0.8930 | 28.62/0.7822 | 28.41/0.7618 | 25.77/0.7760 | 30.70/0.9118 | 30.30/0.8471 |
| | IMDN(CBAM) | 715.4K | 72.080 | 32.18/0.8941 | 28.68/0.7829 | 28.45/0.7627 | 25.84/0.7788 | 30.59/0.9114 | 30.16/0.8461 |
| | **IMDN(BAM)** | 715.1K | 72.021 | 32.24/0.8955 | 28.75/0.7847 | 28.51/0.7642 | 26.08/0.7854 | 31.02/0.9154 | 30.39/0.8492 |

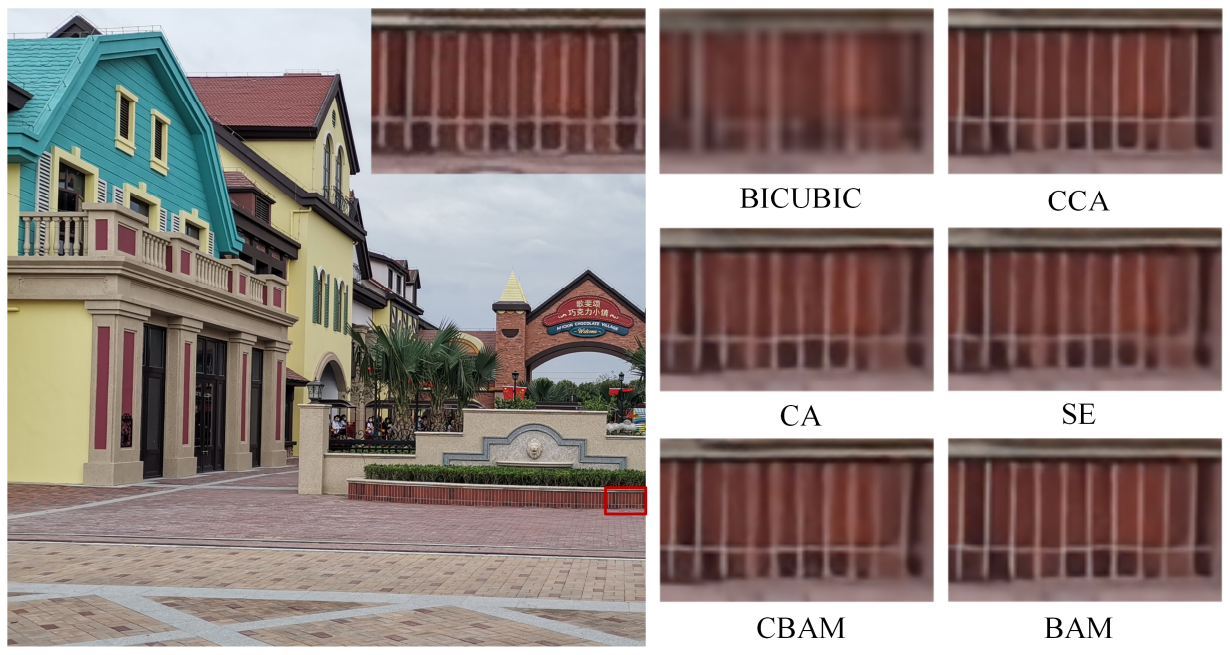

Figure 5: Comparison of ×4 SR results of 5 attention mechanisms on the realSR7 dataset.

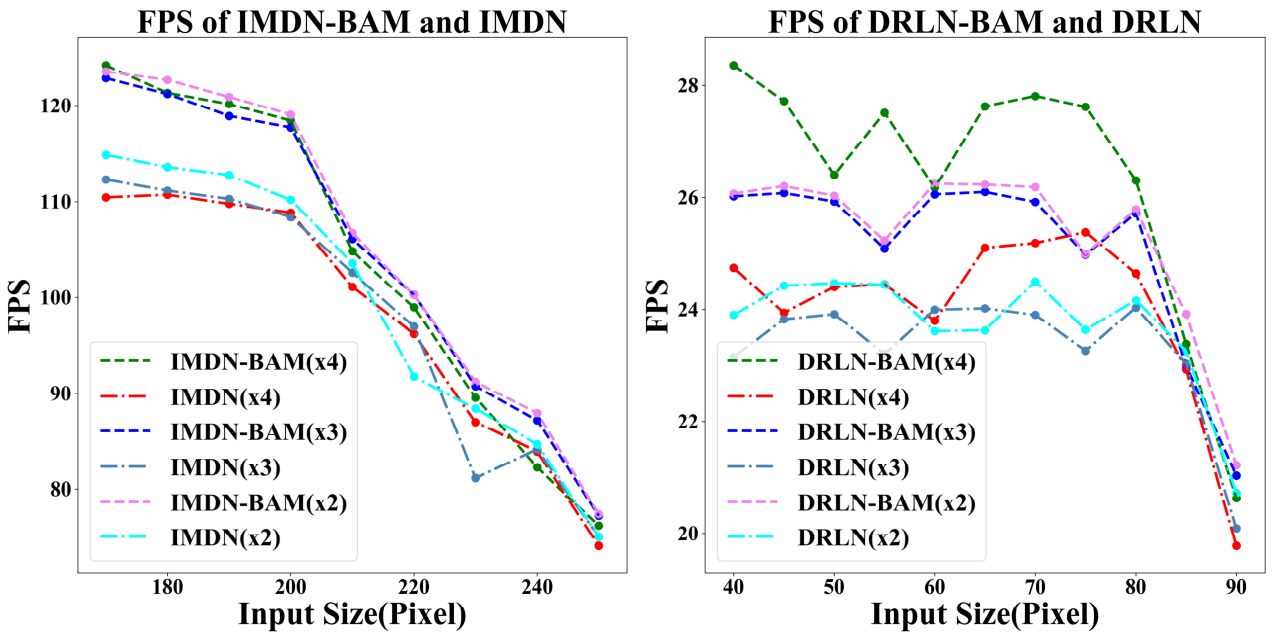

Figure 6: Speed comparison between IMDN-BAM and IMDN, DRLN-BAM and DRLN on 2080Ti.

### Speed comparison

To further prove the efficiency of BAM, we select IMDN and DRLN as the representatives of lightweight and heavyweight SISR networks respectively, and compare the FPS between the experimental group and the control group with different input scales. Under each input scale, we count the average inference time of 700 images to calculate FPS, and it can be expressed as following

$$FPS = Frames \,/\, Time_{Frames}\,, \tag{7}$$

where $Frames$ is the number of images, and $Time_{Frames}$ is the total time utilized for inference. Figure 6 shows the FPS curves of IMDN-BAM and IMDN, DRLN-BAM and DRLN under different input scales on 2080Ti. It can be seen that our BAM has the advantage in inference speed as well, and the speed advantage gets more obvious when the scale of the input image is smaller. When the input image size is 200×200, IMDN-BAM exceeds IMDN {8.9FPS, 9.4FPS, 9.6FPS} under three SR magnifications of ×2, ×3, and ×4 respectively. And when the input image scale is 60×60, DRLN-BAM exceeds DRLN {2.6FPS, 2.1FPS, 2.4FPS} under the three SR magnifications of ×2, ×3, and ×4. The above experimental results illustrate that BAM can accelerate the inference speed while improving network performance indicators, which has significant application value for the landing of lightweight networks on mobile terminals.

## Conclusion

Aiming at the problem that textures are often overwhelmed by extreme noise in SISR tasks, we propose a universal attention mechanism BAM. The overall parallel structure of BAM enables ACAM and MSAM to optimize each other during the back propagation process, so as to obtain an optimal balance between noise suppression and texture restoration. In addition, the parallel structure brings in a faster inference speed. The control experimental results strongly prove that BAM can efficiently improve the performance of SOTA SISR networks and further reduce the parameter amounts and improve the inference speed for those originally with attention. The ablation experimental results illustrate the efficiency of BAM. What's more, BAM demonstrates higher capability to restore the texture aliasing area in real scenes on the realSR7 dataset proposed in this paper.